\documentclass[twocolumn]{aastex63}
\usepackage{amsmath,hyperref,xspace}
\definecolor{RoyalBlue}{HTML}{0071BC}
\definecolor{Blue}{HTML}{2D2F92}
\definecolor{Purple}{HTML}{99479B}
\hypersetup{linkcolor=RoyalBlue,citecolor=Blue,urlcolor=Purple}

\newcommand{\LH}{\textsf{L\!H}\xspace}
\newcommand{\RH}{\textsf{R\!H}\xspace}
\newcommand{\PH}{\textsf{P\!H}\xspace}
\newcommand{\LS}{\textsf{L\!S}\xspace}
\newcommand{\RS}{\textsf{R\!S}\xspace}

\received{2021 July 30}
\revised{2021 December 17}
\accepted{2022 January 24}
\published{2022 February 24: ApJ, 926, 203}

\begin{document}
\vspace*{-5mm}
\title{Radialization of satellite orbits in galaxy mergers}
\shorttitle{Radialization}
\shortauthors{Vasiliev et al.}

\author{Eugene Vasiliev}
\affiliation{Institute of Astronomy, Madingley Road, Cambridge, CB3 0HA, UK}
\affiliation{Lebedev Physical Institute, Leninsky Prospekt 53, Moscow, 119991, Russia}
\author{Vasily Belokurov}
\affiliation{Institute of Astronomy, Madingley Road, Cambridge, CB3 0HA, UK}
\affiliation{Center for Computational Astrophysics, Flatiron Institute, 162 5th Avenue, New York, NY 10010, USA}
\author{N. Wyn Evans}
\affiliation{Institute of Astronomy, Madingley Road, Cambridge, CB3 0HA, UK}

\begin{abstract}
We consider the orbital evolution of satellites in galaxy mergers, focusing on the evolution of eccentricity. Using a large suite of $N$-body simulations, we study the phenomenon of satellite orbital radialization -- a profound increase in the eccentricity of its orbit as it decays under dynamical friction. While radialization is detected in a variety of different setups, it is most efficient in the cases of high satellite mass, not very steep host density profiles, and high initial eccentricity. To understand the origin of this phenomenon, we run additional simulations with various physical factors selectively turned off: satellite mass loss, reflex motion and distortion of the host, etc. We find that all these factors are important for radialization, since it does not occur for point-mass satellites or when the host potential is replaced with an unperturbed initial profile. The analysis of forces and torques acting on both galaxies confirms the major role of self-gravity of both host and satellite in the reduction of orbital angular momentum. The classical Chandrasekhar dynamical friction formula, which accounts only for the forces between the host and the satellite, but not for internal distortions of both galaxies, does not match the evolution of eccentricity observed in $N$-body simulations.
\vspace*{8mm}
\end{abstract}

%%%%%%%%%%%%%%%%%%%%%%
\section{Introduction}

Galaxies experience many mergers with lesser systems in their lifetime. The primary mechanism for the orbital decay of a satellite is dynamical friction, introduced by \citet{Chandrasekhar1943} as a deceleration of a point mass moving through a uniform background of stars. Chandrasekhar's dynamical friction (CDF) acceleration, in the simplified case of a Maxwellian velocity distribution of field stars, is
\begin{equation}  \label{eq:CDF}
\dot{\boldsymbol v} = -\frac{\boldsymbol v}{v} \frac{4\pi\, G^2\, M\, \rho\, \ln\Lambda}{v^2} 
\left[ \mathrm{erf}(X)-\frac{2\,X}{\sqrt{\pi}}\exp(-X^2)\right],
\end{equation}
where $M$ and $v$ are the mass and velocity of the moving body, $\rho$ and $\sigma$ are the density and velocity dispersion of field stars, $\ln\Lambda \sim \mathcal O(1)$ is the Coulomb logarithm, and $X\equiv v/\sqrt{2}\sigma$.
An intuitive physical interpretation of the deceleration as arising from the density wake (gravitational focusing of incoming stars leading to an overdensity behind the moving body) qualitatively agrees with the above formula \citep{Mulder1983}. 

Of course, in reality the host galaxy is not homogeneous, the satellite is not a point mass, and its trajectory is not a straight line. A completely different approach was developed by \citet{Tremaine1984} and follow-up papers, in which the response of the host galaxy to the satellite is computed using linear perturbation theory for a finite, self-gravitating host galaxy, decomposed into appropriately chosen orthogonal modes. In this picture, only the stars whose orbital frequency is in resonance with the satellite contribute to the friction force, regardless of their actual position within the galaxy (i.e., the response is fundamentally global). This perturbative approach breaks down for a sufficiently massive satellite, since its orbit shrinks so rapidly that the notion of resonances becomes useless (though the calculation may still remain qualitatively correct). They argue that a general agreement between the global mode analysis and the manifestly local CDF appears to be fortuitous and should not be expected to hold in every configuration. A counter-example to CDF is provided by a finite-size satellite orbiting just outside the host galaxy (e.g., \citealt{Weinberg1986}): both common sense and numerical experiments suggest that it should excite tides and slow down, but the CDF drag force is zero since the local density is zero.

A number of studies since 1980s, which we review in detail in a later section, explored satellite sinking with various kinds of $N$-body simulations and compared the results to the CDF approximation. The CDF-like ``local response'' of the host galaxy in the form of a density wake behind the moving satellite clearly exists, but appears to be not the only ingredient in the process. The ``global response'' is equally important, although confusion arises over its definition: whether the motion of the host centre is included or not, whether to use the centre of mass or centre of density, etc. Despite obvious limitations, CDF is widely used to estimate the drag force experienced by a sinking satellite, and performs reasonably well with a suitable choice of the Coulomb logarithm (e.g., \citealt{Hashimoto2003}). The evolution of orbital eccentricity is less well reproduced by CDF: the latter typically predicts that the satellite orbits should circularize during sinking, while this effect is far weaker or even reversed in $N$-body simulations. 

\citet{Amorisco2017} studied the orbital properties of satellite debris in galaxy mergers with cosmologically motivated properties, exploring a range of mass ratios and initial orbit parameters. He found that low-mass satellites (mass ratio $\lesssim 1:50$) with initially low-eccentricity orbits tend to circularize, but higher mass or larger initial eccentricity always produced net radialization, as quantified by the final distribution of orbital eccentricities of satellite debris. Although that paper did not examine the physical mechanism of radialization, in a subsequent conference presentation\footnote{https://www2.mpia-hd.mpg.de/homes/stellarhalos2018-loc/sh2018/slides/02.07.Amorisco.pdf}, Amorisco identified the tidal deformation of both galaxies as an important factor in this phenomenon. In retrospect, hints of radialization are already present in the earlier simulations of merging spiral galaxies by \citet[][see especially figures 6 and 7]{Barnes1988}. \citet{Barnes1992} also emphasised the importance of the interaction between each bulge and its own surrounding halo as driving the transfer of binding energy and angular momentum to more outlying parts of the system.

The orbital evolution of a massive satellite and its effect on the host galaxy are relevant for the Milky Way in the context of two recent developments. The first one is a discovery of a strongly radially anisotropic population of stars in the Galactic halo, which is generally interpreted as a product of an ancient (8--10 Gyr ago) merger with a massive satellite galaxy on a highly eccentric orbit \citep{Belokurov2018,Helmi2018}. Using a suite of tailored $N$-body simulations, \citet{Naidu2021} found that a fairly massive (mass ratio $q=1:2.5$) satellite with an initial orbit circularity $\eta=0.5$ is able to reproduce the known properties of the debris very well, and their Figure~8 reveals that its orbit is rapidly radialized during the merger.

The second application is the ongoing encounter of the Milky Way with its largest satellite -- the Large Magellanic Cloud (LMC), whose mass is estimated to be $(1-2)\times10^{11}\,M_\odot$ from various arguments \citep[e.g.,][and references therein]{Erkal2019}, i.e., at least 10\% of the Milky Way mass. The LMC has just recently passed the pericentre of its orbit, likely for the first time \citep[e.g.,][]{Besla2007}, and hence has not yet radialized: its present-day tangential velocity of 310~km\,s$^{-1}$ \citep{Luri2021} at a distance of 50~kpc corresponds to a circularity $\eta=0.7$, using a fiducial Milky Way potential model from \citet{McMillan2017}. However, it causes a significant displacement (reflex motion) of the Milky Way centre with respect to its outer parts \citep{Gomez2015,Petersen2020,Petersen2021,Erkal2021} and a distortion of its halo \citep{GaravitoCamargo2019,GaravitoCamargo2021,Cunningham2020}.

Finally, if the merging galaxies each contain a supermassive black hole (SMBH), then a binary SMBH will eventually form. The efficiency of gravitational-wave radiation, and hence the lifetime of a binary SMBH $\tau$, strongly depends on its eccentricity ($\tau \propto (1-e^2)^{7/2}$, \citealt{Peters1964}), and it is quite plausible that the initial eccentricity at the time of binary formation is largely determined by the orbit of the satellite galaxy, although, to our knowledge, no study has directly addressed this question. Radialization of satellite orbits thus has direct implications for the future gravitational-wave observations.

The growing evidence that massive satellites on initially moderately eccentric orbits undergo further radialization in $N$-body simulations, contrary to the CDF predictions, motivated the present study. We perform a number of $N$-body experiments with different host and satellite density profiles, mass ratios, orbital circularities, and explore the role of several physical ingredients such as the reflex motion of the host and the mass loss by the satellite by selectively turning them off in the simulations. We also analyze the torques acting on each galaxy and their impact on the eccentricity evolution in each setup. We argue that the radialization is a complex phenomenon resulting from several factors, and that the deformability of both the host and the satellite plays an important role.

We first review the previous studies of merging satellites in Section~\ref{sec:literature}. Then in Section~\ref{sec:simulations} we describe our simulation setup, initial condition, and analysis procedures. Section~\ref{sec:results} presents the results and discusses the role of different factors in the radialization process. Finally, Section~\ref{sec:conclusion} wraps up.

%%%%%%%%%%%%%%%%%%%%%%%%
\section{Previous works}   \label{sec:literature}

Several studies in the 1980s tested the validity of CDF against numerical simulations, using various approximations to make the problem computationally tractable.
\citet{Lin1983} introduced a ``semi-restricted $N$-body approach'', in which host particles move in the combined potential of the moving satellite (softened point mass) and a fixed central point mass representing the host galaxy, creating a density wake behind the satellite, while the latter experiences the reciprocal force from the wake and is slowed down. However, in this approach the linear momentum of the system is not conserved, and the self-gravity of host particles is intentionally ignored. On the other hand, \citet{White1983} studied a live self-gravitating $N$-body system, but to solve the Poisson equation without having to perform a costly $\mathcal O(N^2)$ pairwise force computation, he considered several choices for the multipole expansion of the host galaxy potential, using a monopole or quadrupole order and a fixed or moving centre of expansion. \citet{Bontekoe1987} used a similar technique but came to a rather different conclusion from White's study, namely that the self-gravity of the host galaxy is unimportant and a local CDF formula agrees well with the $N$-body results. \citet{Zaritsky1988} examined the source of discrepancy and traced it to the correction step applied by \citet{Bontekoe1987} to restore the conservation of angular momentum, concluding that it was indeed necessary to reach an agreement between different methods, and that the local description seems adequate. 

Nevertheless, the common aspect shared by these codes, namely the low-order multipole approximation of the host galaxy's gravitational force and the need to choose the centre of expansion carefully, still left some concerns about the possible systematic effects. However, \citet{Hernquist1989} soon repeated the simulations with the same initial conditions, but using a tree-code as the Poisson solver, which is free of these deficiencies, and got an excellent agreement with the results of \citet{Bontekoe1987} and \citet{Zaritsky1988} in the self-consistent host galaxy case (all these studies used the same setup, namely an $n=3$ polytrope host profile and a 1:10 merger with a point-like satellite on a circular orbit). They also performed a simulation with a fixed analytic potential of the host, similarly to the method of \citet{Lin1983}, and found the sinking time to be $\sim 2\times$ shorter in the semi-restricted, non-self-consistent case. They argue that this difference corroborates the importance of self-gravity of the host galaxy, at odds with the conclusions of the above mentioned studies. However, in the interpretation of \citet{Hernquist1989}, the self-gravitating response includes the motion of the host galaxy's centre of mass, which is accounted for in the simulations of \citet{Bontekoe1987} and \citet{Zaritsky1988}, but not in the approach of \citet{Lin1983}. In both self-gravitating and non-self-gravitating cases, the results of their $N$-body simulations also agreed reasonably well with the perturbation theory analysis of \citet{Weinberg1986,Weinberg1989}, which involves a global response of the host galaxy.

\citet{Prugniel1992} revisited the problem using self-consistent $N$-body simulations, complemented with various approximate schemes designed to test the role of different physical factors (self-gravity of the host and the satellite, truncation of the multipolar potential approximation at different orders). They concluded that the self-gravity of the wake, i.e., its contribution to the motion of host galaxy particles, is not important, aside from the centre-of-mass motion -- i.e., they followed the convention of \citet{Bontekoe1987} and \citet{Zaritsky1988} rather than that of \citet{Hernquist1989}, arguing that the shift of the host barycentre is not a ``real deformation''. With this convention, the dipole term in the perturbation cancels out, and they find that both quadrupole and octupole perturbations are significant, but higher-order terms in the host potential are less important, thus demonstrating the global character of the response. They also compared the sinking rates of a rigid satellite or a deforming one, and found that the latter loses energy considerably faster, i.e., the self-friction resulting from the satellite deformation is important.

The above mentioned studies mainly considered satellites on circular orbits. \citet{Seguin1996} instead focused on the case of radial or highly eccentric orbits. They found that the sinking rate was not strongly sensitive to eccentricity, and contrary to the previous study, that a deforming and disrupting satellite experiences weaker friction than a point-like one, since self-friction does not make up for the reduction of the wake in the host galaxy. They also stress that the energy loss depends on the entire previous orbit and is not simply determined by the local density of the host galaxy, undermining the validity of the CDF approximation. We note that both this and the previous study used a particle-mesh Poisson solver, and due to a limited spatial resolution, employed various tricks to compensate for numerical errors. It is unclear to what extent their results might be affected by these issues.

\citet{Cora1997} also considered moderately eccentric satellite orbits, again comparing two types of simulations: direct-summation $N$-body code (self-consistent) or a fixed analytic potential of the host galaxy (non-self-consistent, with either a moving or pinned centre); in all cases, the satellite was a Plummer-softened point mass ranging from 0.005 to 0.09, and the host galaxy of unit mass also followed the Plummer profile. As in most previous works, they did not find substantial differences between the evolution of the live host galaxy or a fixed potential with a moving centre, whereas with a pinned centre, the satellite sinking rate was higher. Thus, they concluded that the self-gravity of the host was not important (excluding the centre-of-mass motion), and that the CDF formula adequately described the energy loss rate, except at late stages of evolution of more massive satellites, which affected the structure of the host galaxy considerably. They did not specifically discuss the eccentricity evolution, but from their Figure~2 it seems that the circularity stayed constant or increased.

\citet{vdBosch1999} simulated the orbital decay of point-like, low-mass satellites ($M_\mathrm{sat}/M_\mathrm{host}\lesssim 0.02$) in truncated isothermal host haloes with a tree-code $N$-body integrator. By analyzing the CDF torque at different orbital phases, they conclude that the dynamical friction should reduce the eccentricity near the pericentre and increase around the apocentre. The outcome of the opposite trends might depend on other properties of the system such as the halo density profile, but most studies that use the CDF formula to study the eccentricity evolution find that the satellite orbits tend to circularize, in contrast to $N$-body simulations (e.g., \citealt{Jiang2000,Nipoti2017}). In the $N$-body simulations of \citet{vdBosch1999}, the eccentricity did not appreciably change during the entire evolution. They also found that high-eccentricity orbits decay $1.5-2$ times faster than circular ones.

\citet{Hashimoto2003} examined the discrepancy in the orbital sinking rate and eccentricity evolution between the CDF and $N$-body simulations performed with a hardware-accelerated direct-summation code. They found that by letting the Coulomb logarithm vary with radius, $\ln\Lambda \simeq \ln\big[r(t) / \epsilon \big]$, where $r(t)$ is the instantaneous distance between the satellite and the host centre and $\epsilon$ is the softening length, they were able to match both the decay rate and the eccentricity evolution, since the friction force is somewhat weakened at the pericentre compared to the CDF expression with a constant Coulomb logarithm. In both their $N$-body simulations and the improved CDF approximation, the eccentricity remained roughly constant. This improved CDF formula has been successfully employed in later studies, e.g., \citet{Jethwa2016}, who used it to follow the orbit of the LMC.

Almost all previously described studies used point-like satellites and ignored the tidal stripping. \citet{Fujii2006}, \citet{Fellhauer2007} and more recently \citet{Miller2020} investigated the role of ``self-friction'' -- the orbital decay caused by the material stripped from the satellite itself. All studies found it to be sub-dominant, contributing $\sim10-20\%$ of the total torque, although the latter work also indicates that the retarding torque may be stronger for more eccentric orbits shortly after the pericentre, or even change sign (i.e., increase the angular momentum) around the apocentre.

\citet{Tamfal2021} simulated a 1:10 merger using a modern tree code (\textsc{Pkdgrav3}) with a large number of particles (up to $3\times10^8$). They placed the satellite on a moderately eccentric orbit and considered live (disrupting) satellites and rigid (point-mass) ones. One of the key features in their analysis was the decomposition of the host galaxy density into spherical harmonics, which they used to separate the ``local'' response of the halo (the classical density wake) from the ``global'' modes, primarily the $l=1$ harmonic (dipole). The latter was found to be very prominent in all simulations, appearing right after the first pericentre passage and persisting until the end of the evolution, even when the satellite was already disrupted. They also investigated the torque acting on the satellite (only in the point-mass case) and its dependence on the distance (using only particles within a certain radius to compute the torque). They find that the local wake dominates up until the first pericentre passage, after which the bulk of the torque is created by the global response of the host galaxy. Thus they conclude that the global modes are actually the dominant cause of the orbital decay. Note that in their case, in contrast to Weinberg's analysis, the spherical-harmonic decomposition of the host galaxy is calculated with respect to its (moving) centre of density, thus the dipole perturbation must vanish at small radii and hence the $l=1$ mode represents the internal deformation of the host galaxy rather than its offset from the centre of mass of the entire system (although they do not discuss this explicitly).

To summarize, most of previous work considered simplified cases of point-like satellites and did not specifically address the eccentricity evolution. The CDF approximation adequately describes the energy loss rate, leading to the conclusion by many authors that the local wake is the dominant effect. However, it does not match the increase of eccentricity seen in the cosmologically motivated simulations \citep[e.g.,][]{Amorisco2017}. It seems plausible that the global perturbations in the host galaxy, as well as the self-friction caused by the mass stripped from the satellite, may be important for this radialization~\citep[cf][]{Barnes1988,Barnes1992,Amorisco2017}.

%%%%%%%%%%%%%%%%%%%%%
\section{Simulations}   \label{sec:simulations}

We conduct a suite of collisionless $N$-body simulations of galaxy mergers, using very simple models for both the host and the satellite galaxy: spherical, isotropic, dark matter-only, with prescribed analytic density profiles.

%%%%%%%%%%%
\subsection{Initial conditions}   \label{sec:initial_conditions}

%%%%%%%%%%%%%%
\begin{figure}
\includegraphics{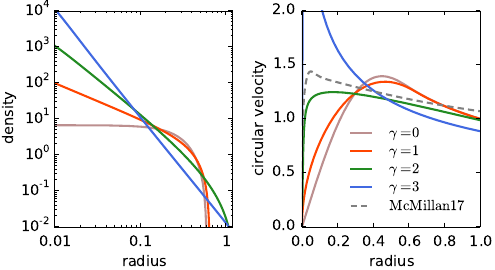}
\caption{Density profiles (left panel) and circular-velocity curves (right panel) of the truncated power-law models considered in this work. They all have the same total mass, but different truncation radii, adjusted so that the circular velocity $v_\mathrm{circ}(r)\equiv \sqrt{M(r)\,r}$ is approximately the same at $r=0.3-0.4$ (roughly the distance of the first pericentre passage for the least eccentric orbit in the series). For comparison, the circular velocity of the Milky Way potential from \citet{McMillan2017} is shown by the dashed grey line, if the models are scaled to physical units as described in the text; the $\gamma=2$ model resembles the Milky Way most closely over a wide range of radii.
}  \label{fig:density_profiles}
\end{figure}
%%%%%%%%%%%%

One goal of this study is to explore the dependence of the galaxy merger on the logarithmic slope of the host density as a function of radius, $\gamma\equiv -\mathrm{d}\,\log\rho/\mathrm{d}\,\log r$. For reference, the standard Navarro--Frenk--White (NFW) profile has a slope $\gamma=1$ for $r\ll r_\mathrm{scale}$ and $\gamma=3$ for $r\gg r_\mathrm{scale}$, although since the profile must be truncated at some finite radius $r_\mathrm{trunc}$ in order to have finite mass, and the concentration parameter $r_\mathrm{trunc}/r_\mathrm{scale}$ rarely exceeds 20, there is actually no extended range of radii where the slope remains constant.  For a cleaner perspective, we decided to construct models with a density as close to a pure power law $(\rho\propto r^{-\gamma})$ as possible; of course, for $\gamma\le 3$, these models must be cut off in radius to have a finite mass, and the density should be capped at radii smaller than the softening length. 

The Eddington inversion formula provides a spherical isotropic distribution function (DF) for the given density profile; however, as discussed in \citet{Lacroix2018}, its application to a density with a sharp cutoff in radius is problematic and leads to a divergent DF at the outer boundary. This divergence could be rectified by imposing a more gentle (e.g., exponential) cutoff, adding a tapering factor $\exp[-(r/r_\mathrm{trunc})^\eta]$ to the density profile. We find that this tapering works adequately for the NFW model or for a $\gamma=3$ power law, but for shallower profiles, the DF still has an unphysical concentration near the outer boundary, and the merger simulations with these models displayed an undesirable propensity to a rapid loss of outer layers of the satellite. Instead, we followed an alternative approach outlined in section 3.1.3 of \citet{Lacroix2018}: we introduce the truncation in the DF itself, not the density profile. 
Specifically, the DF is assumed to coincide with the DF of a pure power law model up to the energy $E=\Phi(r_\mathrm{trunc})$, and set to zero at higher energies. Then the density and the corresponding potential are adjusted to match the truncated DF. This procedure is analogous to the one used by \citet{Woolley1954} to obtain the truncated Maxwellian DF; the more familiar \citet{King1966} model is constructed similarly, but additionally subtracts a constant from the DF so that it reaches zero at the outer boundary, rather than dropping to zero discontinuously (see also \citealt{Gieles2015} for a general case with adjustable smoothness of the outer boundary). We note that even the sharply truncated DF produces a reasonable density profile near the outer boundary: $\rho \propto (r_\mathrm{trunc}-r)^{3/2}$. The cored model ($\gamma=0$) is a polytrope with index $n=3/2$, in which the DF is constant across the entire range of energy. Although we focus on isotropic models in this study, we additionally considered radially and tangentially anisotropic models with the same density profiles, but with the anisotropy coefficient $\beta=0.5$ and $-0.5$, respectively. In these models, the DF is given by %$f(E,L) = -(2\pi^2\,L)^{-1}\,\mathrm{d}(r\rho)/\mathrm{d}\Phi$ for the radially anisotropic model and $f(E,L) = (2\pi^2)^{-1}\,L\,\mathrm{d}^2(\rho/r)/\mathrm{d}\Phi^2$ for the tangentially anisotropic one 
Equations 4.62--4.71 in \citet{Binney2008}.

Figure~\ref{fig:density_profiles} shows the density profiles and circular-velocity curves of the four truncated power-law models we considered ($\gamma$ ranging from 0 to 3). They are all set to have the same total mass of unity (here we use scale-free $N$-body units with $G=1$; a suitable physical scaling to a Milky Way-like galaxy is given by setting the mass unit to $10^{12}\,M_\odot$, length unit to 165~kpc, velocity unit -- to 161~km/s, and time unit -- to 1~Gyr). In order to have a comparable density at the range of radii where the disruption occurs, we adopted different truncation radii (0.6, 0.72, 1.3 and 6.0 for $\gamma=0$ to 3; note that in the latter case, the mass within $r=1$ is $\sim 0.8$, i.e., somewhat lower than in other models, but since the density grows very steeply, it still has a higher enclosed mass within $r=0.4$). The satellite galaxies are chosen to have the same functional form of the density profile as the host, but with scale radii proportional to $\sqrt{q}$, where $q<1$ is the mass ratio between the satellite and the host, hence the velocity dispersion scales as the fourth power of mass, following the observational \citet{Faber1976} relation. We also run simulations with rigid (non-disrupting) satellites as described below. In addition, we also reproduced the simulation setup (initial conditions and trajectories) of several earlier works \citep{Zaritsky1988, Hernquist1989, Cora1997} and obtained a very close agreement with their results.

%%%%%%%%%%%
\subsection{Simulation code and analysis procedures}   \label{sec:nbody_code}

We use the fast-multipole $N$-body simulation code \textsc{gyrfalcON} \citep{Dehnen2000}, chosen for its efficiency and accuracy (the method conserves the linear momentum exactly). To facilitate run-time analysis and various modifications of the $N$-body system during the simulations, we developed a Python interface for the \textsc{falcON} gravity solver and reimplemented the integrator (a simple leapfrog) in Python, together with the additional procedures described in Section~\ref{sec:variants}. The Python module and an example script for running an $N$-body simulation are available at \url{https://github.com/GalacticDynamics-Oxford/pyfalcon}.

The host galaxy is represented by $10^6$ equal-mass particles, and the satellite -- by $0.5\times10^6$, regardless of the mass ratio. We checked that the models are stable when evolved in isolation. The softening length is set to $0.002$ length units (note that \textsc{falcON} uses a custom softening kernel, which approximately corresponds to a Plummer kernel with $1.5\times$ smaller length) and the timestep -- to 0.002 time units for $\gamma\le 2$ models, and four times shorter for the $\gamma=3$ case, which has a very high central density and velocity dispersion. The energy is conserved at a level $\lesssim 10^{-3}$. Doubling the softening length or halving the number of particles did not affect the results.

We start the satellite at $r_\mathrm{init}=1$ with a total velocity (relative to the host) equal to the circular velocity at this radius ($v_\mathrm{init}=0.9$ for the $\gamma=3$ case and 1 otherwise), split in some proportion between the radial and tangential components: $v_\mathrm{tan} = \eta\,v_\mathrm{init}$. Here $0 \le \eta \le 1$ is the circularity parameter defined as
$\eta=L/L_\mathrm{circ}(E)$, the ratio of the instantaneous angular momentum to the angular momentum of a circular orbit with the same energy. It is directly related to the orbit eccentricity $e$ (e.g., for a Kepler potential, $\eta = \sqrt{1-e^2}$), but is easier to measure at any moment during the evolution. We then add a constant offset to the initial position and velocity of all particles in the simulation to ensure that the centre of mass position and velocity of the entire system are zero (except the \PH scenario described below). %and remain so at all times thanks to the momentum conservation guaranteed by \textsc{falcON}).

The simulations are run for 12 time units or until the satellite is fully disrupted, whichever happens earlier. At each timestep, we measure the position and velocity of both the host and the satellite centres by computing the median values over a subset of particles within a radius $r_\mathrm{max}$ from the position of the centre extrapolated from the previous timestep, and then iteratively refining it. This procedure determines the centre of density rather than centre of mass of the whole object (the difference between them is quite important especially at the later stages of disruption); the radius $r_\mathrm{max}$ is set to enclose 10\% of the total mass at the start of the simulation. Once the satellite is close to full disruption, this procedure inevitably breaks down.

In a highly dynamical disrupting system, there is no simple way to determine its bound mass at each moment of time without considering the entire evolution. For instance, the tidal (or Jacobi) radius shrinks drastically near the pericentre passage, so that the enclosed mass may drop precipitously, but then it rebounds as the satellite moves toward the apocentre, and hence the mass increases again, which is clearly not very meaningful. We found that the following procedure produces reasonably behaved, nearly monotonic mass loss histories. In each output snapshot, we determine the bound mass of the satellite by selecting particles with negative total energy, using the relative velocity w.r.t. the satellite centre and the spherically averaged potential of the satellite, computed using only the bound particles themselves, and then iteratively refine the selection. Note that the host potential is not used in this estimate, but fast-moving particles that have been stripped earlier also do not contribute to the bound potential, thus producing a convergent result.

To compute the orbit circularity $\eta$, i.e., to normalize the instantaneous angular momentum of satellite motion relative to the host centre by $L_\mathrm{circ}(E)$, we need the [spherically averaged] gravitational potential in which the satellite moves. This potential represents the host particles, but also should include some or all satellite particles. It sounds natural to include only the unbound satellite debris, but we find that this choice leads to a non-monotonic evolution of circularity around the pericentre passage (first dropping, then rebounding above the initial value, then finally settling back). Thus we opted to include all satellite particles in the calculation of the total host-centered potential (note that due to spherical averaging, bound particles are localized only in radius but not in angles), which eliminated the drop but not the subsequent bump. This may be considered a pragmatic rather than fundamentally motivated choice, given the conceptual difficulty of rigorously defining the energy of the relative motion for two extended and deforming objects. In the end, we cannot trust the fluctuations of circularity around the times of pericentre passages, but at all other times these values are very similar regardless of the way we define the potential.

%%%%%%%%%%%
\subsection{Variants of simulation setup}   \label{sec:variants}

In the baseline simulations, both the host galaxy and the satellite are live $N$-body systems represented with a sufficiently large number of particles; the host galaxy can be deformed and the satellite is disrupting, so this is the only physically realistic setup, which we label as ``live host, live satellite'' (\LH-\LS). However, to isolate the role of various factors and to establish a connection with earlier studies that used both CDF and semi-restricted $N$-body simulations, we perform a number of additional experiments:

\begin{itemize}
\item We replace a live $N$-body satellite galaxy with a single softened particle with analytic density profile $\rho(r) \propto (r^2+\epsilon^2)^{-7/2}$ (identical to the softening kernel used in \textsc{falcON}) and a radius $\epsilon=0.1q^{1/2}$. The satellite particle is not included in the tree construction, since doing so upsets the accuracy of the tree approximation due to large disparity between masses and softening lengths of host particles and the satellite; instead, its mutual forces with host particles are computed directly. This scenario tests the role of the self-friction from the tidally stripped debris, obviously absent in the case of a point-like satellite, and we denote this setup as ``rigid satellite'' (\RS).
\item Independently of the choice between a live and a rigid satellite, we also explore the role of the self-gravity and deformability of the host galaxy, by replacing its potential with the initial analytic profile. Of course, removing the host particles entirely from the simulation would eliminate the dynamical friction itself, so the actual procedure is somewhat more complicated. The force between the host and the satellite particles (or a single particle in the \RS case) is computed with the tree code and applied symmetrically to both particle sets, i.e., the satellite can excite the local density wake and experience the friction from it. The force between satellite particles is also computed self-consistently with a tree code in the \LS case, or is inapplicable in the \RS case. But the force from the host galaxy on its own particles is computed using the initial, unperturbed potential profile, as if the galaxy was not deforming. In doing so, we have a further choice as to where to place the centre of the analytic host galaxy potential. In one case, we fix it at the initial host centre: this ``pinned host'' (\PH) scenario most closely mimics the CDF approximation, especially when coupled with \RS. In the other case, dubbed ``rigid host'' (\RH), we let the host centre move in such a way that the centre of mass of the entire host+satellite system is fixed at the origin.
\item Finally, and again independently of the previous choices, we may also place two identical satellites on diametrally opposite points around the host, so that they orbit symmetrically and the host centre does not move, and also mirror-symmetrize the particles in the host galaxy. In order to avoid the accumulation of force errors that could eventually break the symmetry, we additionally enforced symmetrization of positions and velocities of both host and satellite particles after each timestep. Thus the host galaxy centre always remains fixed in space, and its response to the satellites contains only even-order modes. The main purpose of this experiment, denoted by adding ``2'' to the label, is to explore the role of the reflex motion of the host galaxy, which is eliminated in the two-satellite scenario. We may still opt to compute the host potential self-consistently (\LH-2) or fix it to the initial analytic profile (the cases \PH-2 and \RH-2 are equivalent).
\end{itemize}

%%%%%%%%%%%%%%%
\begin{figure*}
\includegraphics{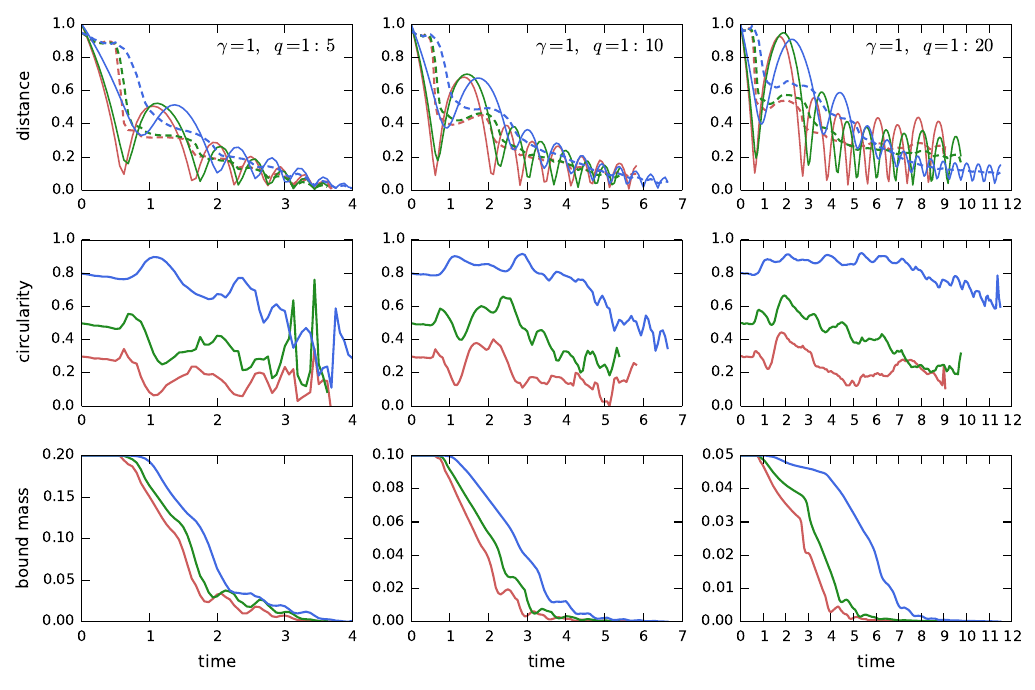}
\caption{Evolution of satellite orbits in the $\gamma=1$ case. From top to bottom: distance between the host and the satellite centres (solid lines) and the semimajor axis of the satellite orbit (dashed lines); circularity $\eta \equiv L/L_\mathrm{circ}(E)$; bound mass. Different columns show mass ratios 1:5 (left), 1:10 (centre) and 1:20 (right), and different colours -- initial circularity 0.8 (blue), 0.5 (green) and 0.3 (red).
}  \label{fig:evolution_gamma1}
\end{figure*}

\begin{figure*}
\includegraphics{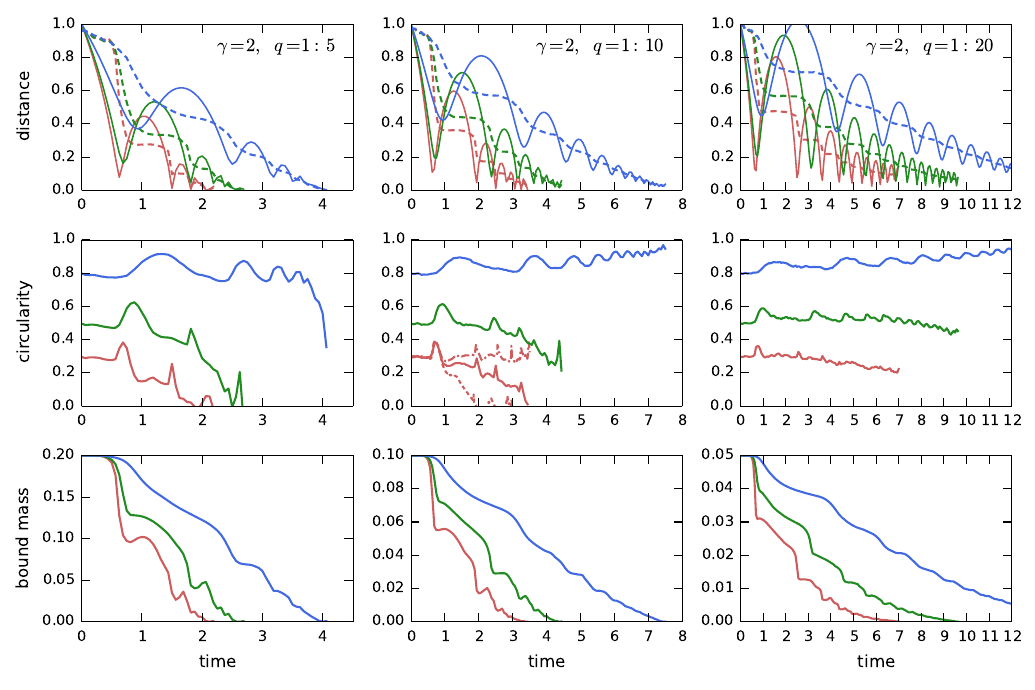}
\caption{Same as Figure~\ref{fig:evolution_gamma1}, but for the case $\gamma=2$. In the central panel, we additionally show the evolution of circularity in the two simulations with anisotropic host galaxies (dotted -- radial, dot-dashed -- tangential).
}  \label{fig:evolution_gamma2}
\end{figure*}

\begin{figure*}
\includegraphics{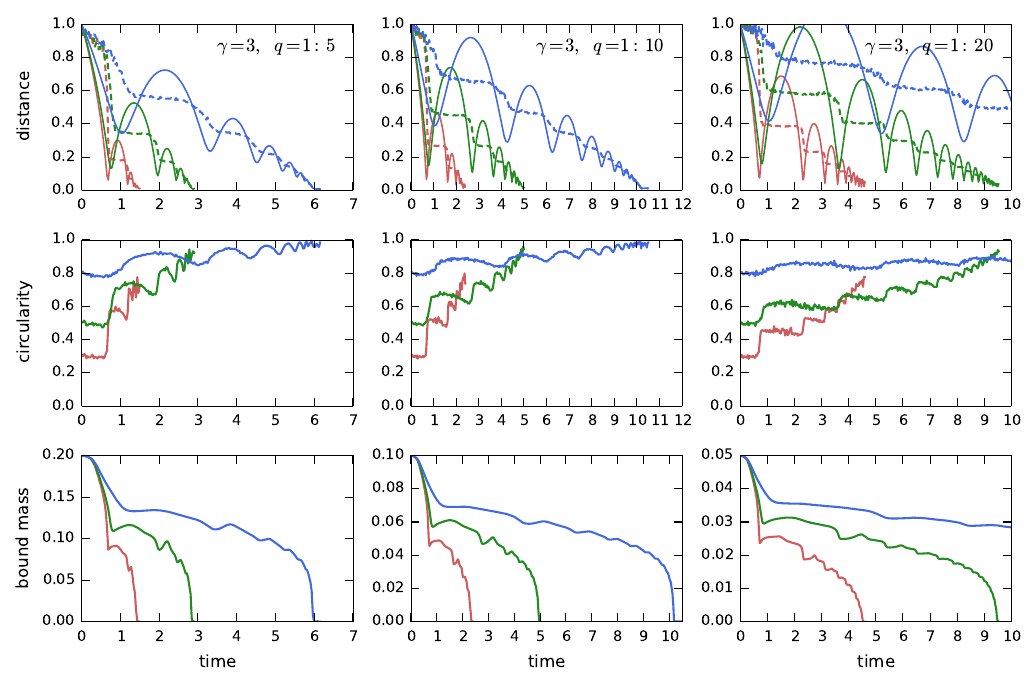}
\caption{Same as Figure~\ref{fig:evolution_gamma1}, but for the case $\gamma=3$.
}  \label{fig:evolution_gamma3}
\end{figure*}
%%%%%%%%%%%%%

As described earlier, most of the previous studies focused on the \RS scenario, thus comparing it with \LS allows us to establish the role of self-friction from the stripped satellite debris. A complicating factor is that a disrupting satellite experiences less friction at later stages because of a much smaller remaining mass; thus we cannot directly compare the evolution of \RS and \LS with the same initial mass. We find that using a twice lower initial mass for \RS comes closest to being a fair comparison.

The comparison between \LH and \RH or \PH probes the role of the host deformation, which may or may not include the centre-of-mass motion. The \PH-\RS case is equivalent to the ``semi-restricted'' $N$-body approach of \citet{Lin1983}, and a number of subsequent studies \citep{White1983, Bontekoe1987, Cora1997} concluded that the satellite orbit decays faster in this case, because the density wake in the host galaxy has a dipolar pattern if considered with respect to the initial host centre. By contrast, in the \RH case, the host centre moves around in such a way as to compensate the dipole perturbation in the central parts, although it does not cancel it exactly because the centre of mass and centre of density do not coincide, as discussed later. Moreover, the deformation of the host is not uniform in space: the inner parts may be able to respond faster to the moving satellite than the outer parts, thus a dipolar pattern at large radii will still appear even in the \RH setup. A comparison of \RH with \PH tests the role of the reflex motion, and a comparison of \RH with \LH\ tests the deformation of the outer halo, and hence genuinely the self-gravity of the host. In the two-satellite setup, inspired by \citet{Hernquist1989}, the reflex motion and the dipole are absent by construction, thus \RH-2 vs. \LH-2 tests the role of even-order perturbations in the host.

%%%%%%%%%%%%%%%%%
\section{Results}   \label{sec:results}

%%%%%%%%%%%
\subsection{Baseline simulations}   \label{sec:baseline}

We first consider the only physically realistic setup, that of a live host and live satellite (\LH-\LS). 
Figures~\ref{fig:evolution_gamma1} to \ref{fig:evolution_gamma3} show the evolution of satellite orbit parameters and the bound mass in the series of simulations with $1:20\le q \le 1:5$ and $\gamma=1,2,3$. We considered but excluded from further analysis the simulations with higher mass ratios (e.g., $q=1:2$), since the merger happens too quickly (within $1-2$ orbits), and the $\gamma=0$ case, because the live satellites get rapidly disrupted after the first pericentre passage, precluding any meaningful estimate of their orbital evolution. In the remaining cases, we observe clear trends with the mass ratio $q$, initial circularity $\eta$, and the density slope $\gamma$:
\begin{itemize}
\item Disruption times increase for lower mass satellites. This is not unexpected, since the dynamical friction efficiency is proportional to the satellite mass, although the latter also evolves in the process, leading to a sub-linear scaling of the disruption time with $1/q$.
\item Higher initial circularity leads to a longer lifetime, but this effect also strongly varies with $\gamma$. In the shallow cusp case ($\gamma=1$), there is little difference between the decay histories of highly and moderately eccentric orbits, whereas for the strongly centrally concentrated systems ($\gamma=3$), the tidal forces grow much faster as the satellite approaches small radii, so the highly eccentric orbits lose energy more rapidly, and the orbital period also quickly decreases as the satellite sinks deeper, accelerating its destruction in a runaway manner.
\item Most interestingly, the circularity systematically decreases with time in some cases and increases in other. In the $\gamma=1$ system, almost all satellite orbits exhibit radialization, though with some noise and occasional non-monotonic evolution. In the $\gamma=2$ case, the radialization occurs for higher-mass satellites or for low enough initial circularity, while the orbits rather tend to circularize if started at $\eta=0.8$ and $q\le 0.1$. Finally, in the $\gamma=3$ case, all orbits circularize, especially rapidly if they were initially more radial.
\item Although we focus on isotropic models in the rest of this paper, we report preliminary results of two simulations of a $\gamma=2,\, q=1\!:\!10,\, \eta=0.3$ merger with radially or tangentially anisotropic host galaxies, shown in the central panel of Figure~\ref{fig:evolution_gamma2}. It appears that radial or tangential anisotropy of the host respectively enhances or inhibits the radialization of the satellite, but a detailed analysis of this phenomenon is beyond the scope of this study.
\end{itemize}

The effect of radialization appearing in some of these simulations and its underlying physical mechanisms are worth a deeper investigation. A given amount of deceleration $\Delta v$ reduces the angular momentum more than the energy near the apocentre (hence acts to increase eccentricity), and conversely, reduces energy much more dramatically than the angular momentum near the pericentre (hence decreasing eccentricity). Averaging the CDF force (Equation~\ref{eq:CDF}) over the orbit in a power-law background potential, we find that for any density slope $\gamma>0$, the net result is circularization. On the other hand, \citet{Gould2003} examined the eccentricity evolution in a power-law stellar cusp, but with the total potential dominated by a more massive central black hole. In this case the phase-space DF of background particles drops toward the centre if the density slope $\gamma<3/2$, producing too little friction near the pericentre and thus radializing the orbits, while steeper cusps still lead to circularization. We conjecture that in any self-consistent stellar system with a decreasing density profile, which has a monotonically decreasing DF with energy, the CDF force circularizes orbits. Naturally, this effect would be stronger in steeper density profiles (or, taken to the extreme, in the ``grazing encounter'' scenario of \citealt{Bontekoe1987}, where the density is nonzero only during pericentre passages). \citet{vdBosch1999}, performing numerical simulations of sinking satellites rather than relying on CDF, find that for their host galaxies, modelled as truncated isothermal ($\gamma=2$) profiles, the gain and loss of eccentricity nearly balances along the orbit, but they still see no indication of radialization. However, they considered much smaller satellite masses ($q \le 0.02$) and modelled them as single softened particles, hence neglecting mass loss.

%%%%%%%%%%%
\subsection{Additional experiments}  \label{sec:tweaks}

%%%%%%%%%%%%%%
\begin{figure}
\includegraphics{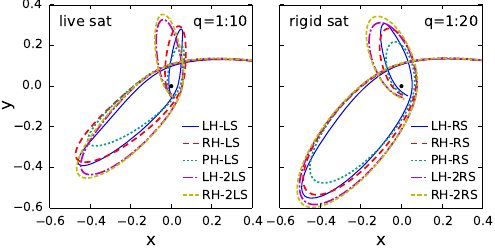}
\caption{Orbits for the case $\gamma=2$, $\eta=0.3$ and either a live satellite with $q=1:10$ (left panel) or a point-like satellite with $q=1:20$ (right panel). Different lines represent variants of the simulation setup as described in Section~\ref{sec:variants}: live, rigid, or pinned host and one or two satellites. In all cases, the orbits are shown in the reference frame centered on the host centre at all times, so that the host is fixed at origin (shown by the black dot). The radialization of the orbit is most prominent in the \LH-\LS case (left panel, blue solid line).
}  \label{fig:orbits_gamma2}
\end{figure}

\begin{figure}
\includegraphics{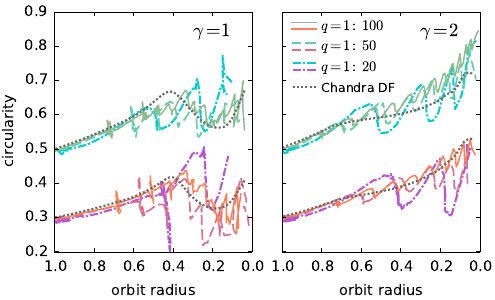}
\caption{Evolution of circularity $\eta$ in $\gamma=1$ (left panel) and $\gamma=2$ (right panel) host galaxies with a compact, non-disrupting satellite (\LH-\RS) of mass $1:100$ (solid), $1:50$ (dashed) and $1:20$ (dot-dashed lines) and initial circularity 0.5 (upper/green curves) or 0.3 (lower/red curves). Instead of time evolution, we plot $\eta$ as a function of the orbit extent (radius of a circular orbit with the same energy, which would be a semimajor axis in the Kepler potential), which changes from 1 (initial value) to zero by the end of the simulation, when the satellite sinks to the centre of the host. For comparison, we also plot the evolutionary tracks computed using the classical Chandrasekhar dynamical friction formula (dotted lines), with a distance-dependent Coulomb logarithm as per \citet{Hashimoto2003} prescription: $\ln\Lambda = \ln [r(t)/\epsilon]$, where $\epsilon$ is the size of the satellite and $r(t)$ is the instantaneous distance from the host centre. Both $N$-body simulations and analytic curves do not show a tendency of radialization characteristic of the cases of extended and disrupting satellites shown in Figures~\ref{fig:evolution_gamma1} and \ref{fig:evolution_gamma2}.
}  \label{fig:evolution_pointmass}
\end{figure}

\begin{figure}
\includegraphics{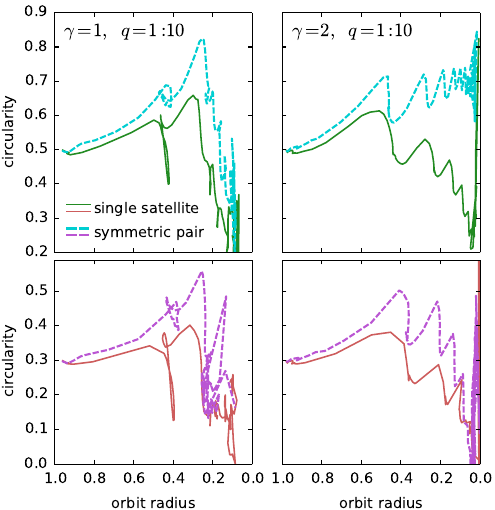}
\caption{Evolution of circularity $\eta$ in $\gamma=1$ (left panels) and $\gamma=2$ (right panels) systems, comparing the baseline simulations of a single, disrupting satellite and moving host galaxy (solid lines, same simulations as shown in the middle column of Figures~\ref{fig:evolution_gamma1} and \ref{fig:evolution_gamma2} respectively) with an artificial setup in which two identical satellites are placed on opposite sides of the host galaxy and evolve synchronously (dashed lines), hence the host remains stationary. Upper panels show the initial circularity of 0.5 and lower -- 0.3. As in Figure~\ref{fig:evolution_pointmass}, we plot $\eta$ as a function of the orbit extent, which changes from 1 (initial value) to 0 by the end of the simulation, when the satellite sinks to the centre of the host. The symmetrically placed satellites do not radialize as much as in the baseline scenario, suggesting that the odd-order modes in the host galaxy (e.g., its displacement from origin) play an important role in this process.
}  \label{fig:evolution_pair}
\end{figure}
%%%%%%%%%%%%

There are at least two possibly important physical ingredients in our simulations that could be responsible for radialization. One is the motion and deformation of the host galaxy in response to the gravitational pull of the satellite. The other is the mass loss from the satellite: the stripped material is distributed asymmetrically and exerts force on the satellite itself. We now turn to the analysis of other simulation setups besides \LH-\LS, which are intended to qualitatively examine the individual role of these factors, neglecting their possible interplay in the actual physically relevant scenario.

To begin with, we consider a fiducial case of a $\gamma=2$ host galaxy and an initially rather eccentric orbit with circularity $\eta=0.3$, comparing ten different simulation setups: disrupting or point-like satellite, live or rigid host, one or two symmetrically placed satellites. We use mass ratio $q=1:10$ for the live satellite and twice lower for the rigid one (which roughly matches the bound mass of the live satellite after the first pericentre passage). Figure~\ref{fig:orbits_gamma2} shows the corresponding orbits up to the third pericentre passage, demonstrating that the most realistic scenario (\LH-\LS) leads to the strongest radialization, and any deviation from this baseline setup diminishes or reverses the trend. In particular, a comparison of \LH (blue solid) and \RH (red dashed) cases demonstrates that the energy evolution is roughly identical (the orbit sizes are similar), but the eccentricity is lower for \RH. Thus we conclude that the deformation of the host galaxy reduces the orbital angular momentum -- a point to which we will return later. The \PH case differs from \RH in that the host potential is fixed in space, preventing not only the deformation, but also the centre-of-mass motion of the host. The corresponding orbits (cyan dotted) decay significantly faster, in agreement with earlier studies that used different host profiles and nearly circular satellite orbits \citep{White1983,Zaritsky1988,Hernquist1989}. In the case of two symmetrically placed satellites, there is little difference between live (purple dot-dashed) and rigid (yellow dashed) hosts, demonstrating that the even-order modes in the host galaxy are relatively unimportant; in both cases the orbits also remain more circular than in the single-satellite setup. Finally, point-like (rigid) satellites demonstrate comparatively less radialization in all cases, although the comparison is complicated by the gradual mass loss of a live satellite.

We also explored these different simulation setups for other values of $\gamma$, mass ratio and initial circularity, focusing on the evolution of circularity $\eta$ plotted against the orbit size (radius of a circular orbit with the given energy) instead of time, to facilitate the comparison between different mass ratios. Figure~\ref{fig:evolution_pointmass} shows $\eta$ for the \LH-\RS case with $\gamma=1,2$ and mass ratios $q=1:20$, $1:50$ and $1:100$. More massive  satellites appear to retain somewhat lower circularity during the simulation, but for all mass ratios, we did not observe as clear radialization as in the disrupting case: if anything, the orbits tend to become more circular, especially in the $\gamma=2$ case. We thus conclude that the mass lost from a disrupting satellite is an important factor in the radialization of its orbit, although the two series of simulations are not directly comparable, since a disrupting satellite sinks more gradually than a compact one.

We compared the evolution of circularity in the \RS simulations with the CDF approximation. The only free parameter in Equation~\ref{eq:CDF} is the Coulomb logarithm $\ln\Lambda$, and it is customary to calibrate its value so that the orbits best match the actual $N$-body simulations. We find that when assuming a constant value, the CDF orbits circularize in all cuspy profiles, since the stronger friction at pericentre preferentially reduces energy, in line with earlier studies. \citet{Hashimoto2003} performed a similar experiment with point-mass satellites, and suggested that a distance-dependent Coulomb logarithm, $\ln\Lambda \simeq r(t) / \epsilon$, where $\epsilon$ is the softening length of the satellite particle and $r(t)$ is its instantaneous distance from the host centre, is able to reproduce the $N$-body results much better than a constant $\ln\Lambda$, both in terms of energy loss rate and eccentricity evolution. With this modification, the friction is somewhat suppressed at pericentre, thus avoiding circularization. We confirm their result: using the same prescription and setting $\epsilon$ equal to the truncation radius of the satellite, we obtain a quantitative agreement between $N$-body and CDF evolutionary tracks. The $\gamma=1$ host system is more compact (its radius is $0.65$), so the satellite apocentre is initially outside the host galaxy and hence it circularizes in the regime of grazing encounters, but after reaching the inner, power-law part of the host density profile, it exhibits a mild drop in circularity. Nevertheless, even this drop is much weaker than found in the simulations of disrupting satellites. In the $\gamma=2$ case, the circularity increases monotonically, though slower for higher-mass satellites. 

Figure~\ref{fig:evolution_pair} compares the evolution of circularity between the single and two symmetric satellite cases (\LH-\LS and \LH-2\LS) with $\gamma=1,2$ and $q=1:10$. As discussed above, the somewhat artificial placement of two symmetric satellites eliminates the centre of mass motion, and the host galaxy develops only even-order perturbations -- the classical trailing wakes as in the CDF picture. In this case, we again see that the 2\LS orbits tend to stay more circular than in the baseline scenario of a single satellite, though the circularity dropped toward the end of the simulation (except the case $\gamma=2$, $\eta=0.5$). We caution that the sinking satellite pairs are not fully equivalent counterparts of single satellites, since the stripped debris have a different spatial distribution relative to the satellite in these two cases.

To summarize, it appears that both the stripping of the satellite and the motion of the host centre are important ingredients in the radialization process. To better understand the physical mechanisms of these factors, we consider the forces acting on the host and the satellite during their evolution.

%%%%%%%%%%%
\subsection{Torques and deformations of the host galaxy}  \label{sec:torques}

%%%%%%%%%%%%%%%
\begin{figure*}
\includegraphics{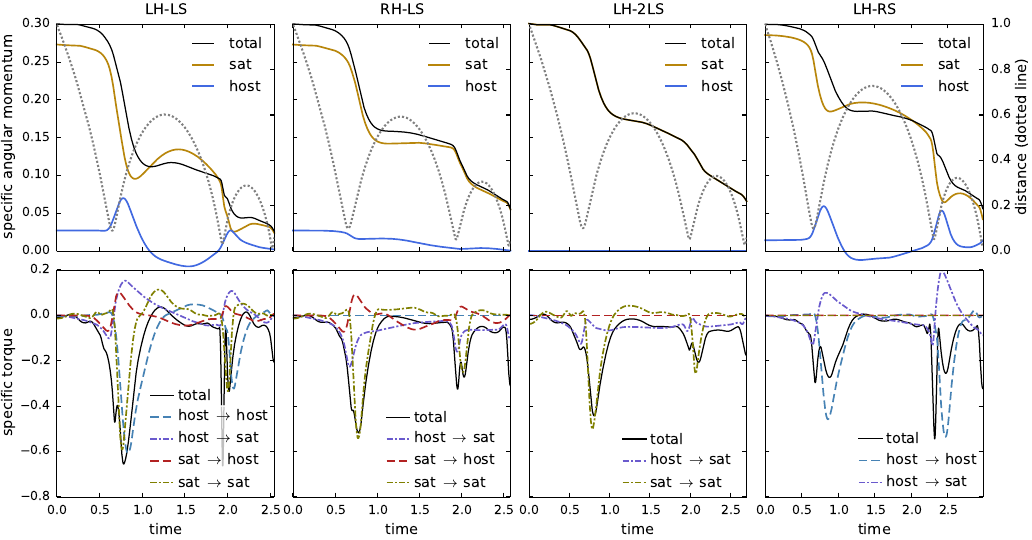}
\caption{
Decomposition of torques in simulations with different setups. The first column is the baseline case (live host and satellite), and remaining columns examine the effect of different physical ingredients: replacing the host potential by an unperturbed analytic one, but still with a moving centre (second column), cancelling the host centre motion by placing two symmetric satellites (third column), and eliminating the tidal stripping by using a point-like satellite (last column). In all cases, the host galaxy is a $\gamma=2$ model, the initial circularity is $\eta=0.3$, and the initial mass ratio is $q=1:10$ for the disrupting satellite or $q=1:20$ for a point-like one. Top row shows the evolution of specific angular momentum of the satellite $\boldsymbol{L}_\mathrm{s}$ (top, yellow), host $\boldsymbol{L}_\mathrm{h}$ (bottom, blue -- in the symmetric case it is identically zero), and their sum (total orbital momentum $\boldsymbol L$, black); dotted line shows the evolution of separation between the host and satellite centres $|\Delta\boldsymbol{x}|$. Bottom row shows the measured torques $\Delta \boldsymbol x \times \boldsymbol a^\mathrm{from}_\mathrm{to}$, where the subscript of the acceleration indicates the point at which it is measured (the centres of either galaxy), and the superscript -- particles that create the acceleration: cyan dashed -- host galaxy acting on itself ($\boldsymbol a_\mathrm{h}^\mathrm{h}$), violet dot-dashed -- host galaxy acting on the satellite ($\boldsymbol a_\mathrm{s}^\mathrm{h}$), red dashed -- satellite acting on the host ($\boldsymbol a_\mathrm{h}^\mathrm{s}$), olive dot-dashed -- satellite on itself ($\boldsymbol a_\mathrm{s}^\mathrm{s}$), and black line is the total time derivative $\mathrm{d}\boldsymbol L/\mathrm{d}t$. We see, counterintuitively, that in the first (physically realistic) case, the self-gravity of both galaxies, creating a negative torque after the pericentre passages, is the dominant reason for the decrease of orbital angular momentum.
}  \label{fig:torques}
\end{figure*}

\begin{figure*}
\includegraphics{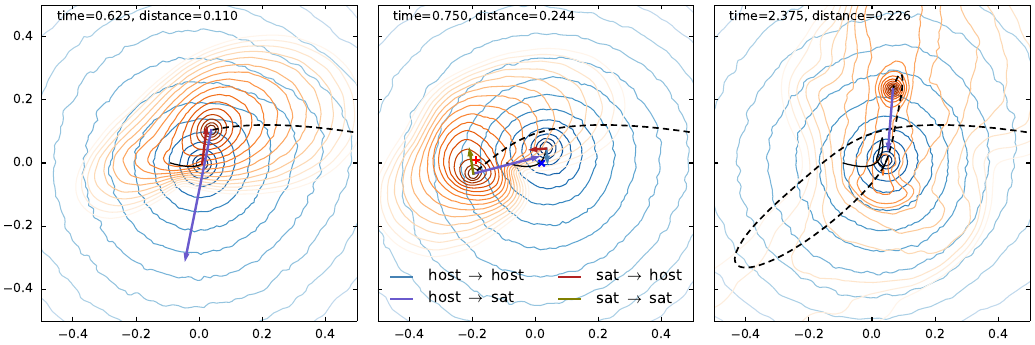}
\caption{Snapshots from the merger simulation with $\gamma=2$, $q=1:10$ and initial circularity $\eta=0.3$. Blue and orange contours show the projected density of the host and the satellite, respectively; their past orbits in the centre-of-mass reference frame are shown by black solid and black dashed lines; and coloured arrows show the forces acting on each galaxy's centre from its own particles and from the other galaxy. In the first snapshot, just before the first pericentre passage, the mutual forces (red and violet arrows) are directed almost along the line connecting the galaxy centres, with a small tangential component responsible for negative torques (decrease in angular momentum due to dynamical friction from the density wake behind the satellite). In the second snapshot, after the pericentre passage, the tangential component of mutual forces create positive torques, but a much larger negative torque is maintained by self-force acting on each galaxy's centre from its own asymmetric mass distribution (cyan and yellow arrows). The self-force of the host galaxy, in particular, arises from the swinging motion of its centre toward the satellite's pericentre point, which displaces the central region relative to the outer parts of the galaxy (the host and satellite centres of mass are marked by blue and red crosses, respectively, and are noticeably offset from their density centres). This recoiling motion is halted and reversed by the gravity of the outer regions, and by the time of the second apocentre passage of the satellite, shown in the third snapshot, the host galaxy centre is back near the origin, and the orbit of the satellite became much more radial.
}  \label{fig:snapshots_gamma2}
\end{figure*}

\begin{figure*}
\includegraphics{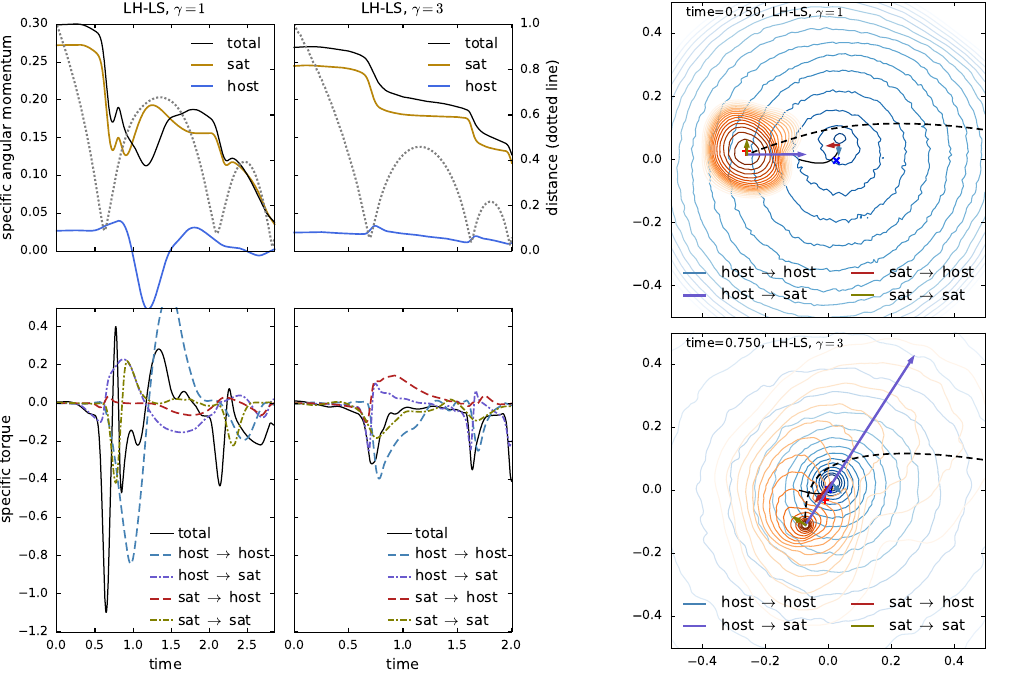}
\caption{Same as Figures~\ref{fig:torques} and \ref{fig:snapshots_gamma2}, but for the simulations with $\gamma=1$ (left column and top right panel) and $\gamma=3$ (centre column and bottom right panel), $q=1:10$ and initial circularity $\eta=0.3$.
}  \label{fig:torques_snapshots_gamma13}
\end{figure*}
%%%%%%%%%%%%%

The evolution of orbit circularity is determined by the interplay between energy and angular momentum losses by the satellite, but the measurement and interpretation of these quantities are not straightforward. It is natural, at least from the observational perspective, to use the potential of the host galaxy and the relative velocity of the satellite with respect to the host galaxy's centre of density (which is different from its centre of mass). However, this reference frame is manifestly non-inertial, complicating the interpretation of the total energy of the satellite. We thus focus on the angular momentum evolution in the first place.

Denote the positions, velocities and accelerations of the host and the satellite centres in the barycentric inertial frame as $\boldsymbol{x}_\mathrm{h}, \boldsymbol{x}_\mathrm{s}$, $\boldsymbol{v}_\mathrm{h}, \boldsymbol{v}_\mathrm{s}$, and $\boldsymbol{a}_\mathrm{h}, \boldsymbol{a}_\mathrm{s}$, respectively. Here the positions of galaxy centres of density are determined iteratively from the coordinates of particles in a small region around the corresponding galaxy's centre, as explained in Section~\ref{sec:nbody_code}, and the velocities and accelerations are defined as first and second time derivatives of the position vectors. Alternatively, we may measure the velocity and acceleration of the centre in the same way as the position (averaging over the same set of particles in the central region); this allows us to decompose the acceleration of each galaxy's centre into the contributions from all host particles and all satellite particles separately: $\boldsymbol{a}_\mathrm{h}^\mathrm{h}$ is the acceleration of the host centre due to the asymmetric mass distribution of the host, $\boldsymbol{a}_\mathrm{h}^\mathrm{s}$ is the acceleration of the host centre due to the gravity of the satellite, etc. We stress that since the centre is a fictitious point, the mean velocity and acceleration of particles in this region are not identical to the time derivatives of the centre position (in other words, $\boldsymbol{a}_\mathrm{h} \ne \boldsymbol{a}_\mathrm{h}^\mathrm{h} + \boldsymbol{a}_\mathrm{h}^\mathrm{s}$).

The position of the satellite in the host-centered reference frame is $\Delta\boldsymbol{x} \equiv \boldsymbol{x}_\mathrm{s}-\boldsymbol{x}_\mathrm{h}$, its specific angular momentum is $\boldsymbol{L} = \Delta\boldsymbol{x} \times (\boldsymbol{v}_\mathrm{s}-\boldsymbol{v}_\mathrm{h})$, and its time derivative is the specific torque $\boldsymbol{T} = \Delta\boldsymbol{x} \times (\boldsymbol{a}_\mathrm{s}-\boldsymbol{a}_\mathrm{h})$. We stress again that these quantities are strictly defined only for point masses, whereas we apply them to fictitious density centres of extended structures. For visualization purposes, we represent $\boldsymbol{L}$ as a sum of contributions of the host and the satellite: $\boldsymbol{L}_\mathrm{h} \equiv -\Delta\boldsymbol{x} \times \boldsymbol{v}_\mathrm{h}$, $\boldsymbol{L}_\mathrm{s} \equiv \Delta\boldsymbol{x} \times \boldsymbol{v}_\mathrm{s}$. Assuming that the total accelerations of both centres are approximately given by the vector sum of accelerations induced by each galaxy's particles, we decompose the torque into four terms, $\Delta\boldsymbol{x} \times \boldsymbol{a}_\mathrm{\{h,s\}}^\mathrm{\{h,s\}}$. Thus the rate of change of angular momentum is $\mathrm{d}\boldsymbol{L}_\mathrm{h}/\mathrm{d}t = -\Delta\boldsymbol{x}\times(\boldsymbol{a}_\mathrm{h}^\mathrm{h}+\boldsymbol{a}_\mathrm{h}^\mathrm{s}) - \boldsymbol{v}_\mathrm{s}\times\boldsymbol{v}_\mathrm{h}$, $\mathrm{d}\boldsymbol{L}_\mathrm{s}/\mathrm{d}t = \Delta\boldsymbol{x}\times(\boldsymbol{a}_\mathrm{s}^\mathrm{h}+\boldsymbol{a}_\mathrm{s}^\mathrm{s}) + \boldsymbol{v}_\mathrm{s}\times\boldsymbol{v}_\mathrm{h}$ (note that the cross-product of velocities is typically small but nonzero, since they are not exactly antialigned at all times, and cancels in the total torque). Because the orbit of the satellite lies in a single plane ($xy$), it is sufficient to consider only the $z$ component of the angular momentum and the torque.

Figure~\ref{fig:torques}, top row, shows the time evolution of the specific angular momentum $L$ (i.e., the distance between the galaxy centres times the relative tangential velocity) for the case of a $\gamma=2$ host and initial circularity $\eta=0.3$ (same configuration as in Figure~\ref{fig:orbits_gamma2}). We compare four scenarios: \LH-\LS, \RH-\LS, \LH-2\LS, and \LH-\RS, examining the role of different physical ingredients. The apocentre radius of the orbit after two pericentre passages is similar among all four cases, but the angular momentum drops much faster in the first (physically realistic) case, i.e., the orbit radializes, unlike in the other setups.

We first draw attention to the fact that the angular momentum drops precipitously \textit{after} each pericentre passage, not \textit{during} it. In the CDF approximation, we might expect that the friction force, which depends mainly on the host density (i.e. on the distance to the host centre),  would be roughly symmetric on both approaching and receding orbital segments, but the actual evolution is in clear contradiction with this simple picture, as already pointed out by \citet{Seguin1996}. Moreover, in the cases of a live host (first and last panels), there is a significant temporal variation of the host galaxy's contribution $\boldsymbol{L}_\mathrm{h}$ to the total angular momentum (blue curves). 
Initially, the ratio of velocities of the host and the satellite galaxies is $q$, i.e., inversely proportional to their masses. As the satellite is passing its first pericentre, it starts to lose angular momentum, but at the same time, the entire central region of the host galaxy centre swings toward the satellite, acquiring a significant tangential velocity (distinct bumps on the blue curves). This swinging motion displaces the host galaxy centre of density relative to its centre of mass (i.e., the outer regions do not experience this rapid displacement), and subsequently the host centre rebounds back, reducing its angular momentum even to negative values (i.e. its velocity is directed in the same sense as the satellite's velocity when the latter reaches its apocentre).
This effect is absent by construction in the third panel (symmetric placement of satellites keeps the host centre fixed at origin), and is also artificially eliminated in the case of a rigid host potential (second panel), whose centre exactly mirrors the motion of the satellite scaled by the instantaneous mass ratio (thus is nearly absent at late times, when the satellite lost most of its mass).

The bottom row of Figure~\ref{fig:torques} shows the decomposition of torques in the same four simulations. We track separately the force from host and satellite particles acting on each galaxy's centre, i.e., there are four components in the total torque (though some of them may be identically zero in non-physical scenarios: a point-like satellite does not exert any torque, and the self-force of host particles is ignored in the \RH scenario). We also plot the time derivative of the specific angular momentum $\boldsymbol T = \mathrm{d}\boldsymbol L/\mathrm{d}t$. As explained above, since the time derivative of relative velocity does not exactly match the sum of accelerations from both galaxies, the derivative of angular momentum is not identical to the sum of the four components, but is close enough at almost all times. Despite these caveats, it is still worthwhile to examine the temporal evolution of different contributions to the torque to gain qualitative understanding of the most important physical ingredients.

In the physically realistic \LH-\LS scenario (first column), we see, unexpectedly, that the main contributions to the total torque come from self-gravity of both the host and the satellite (cyan and yellow curves), i.e., forces created by asymmetries in their respective mass distributions. What is even more surprising is that the other two components (mutual torques, shown by purple and red curves) actually change sign shortly after the pericentre passage, i.e., act to increase the angular momentum, though are subdominant to the first two components. We argue that this counter-intuitive behaviour stems from the swinging motion of the host galaxy centre. 

Figure~\ref{fig:snapshots_gamma2} illustrates the direction of forces and corresponding torques at three moments in the \LS-\LH simulation: just before the first pericentre passage, shortly after it, and after the second apocentre; the colours are the same as in the previous figure. The motion of both host and satellite centres are counter-clockwise; if the force direction is counter-clockwise from the line connecting the host and the satellite centres, it creates a negative torque and reduces the angular momentum, and vice versa. In the first snapshot, the forces on both the host and the satellite are almost entirely caused by the other galaxy and provide slightly negative torques. However, in the second snapshot, there is a clear offset of both galaxies' centres of density from their centres of mass (marked by crosses), and the directions of self-forces (blue and yellow) reflect this offset and create strongly negative torques, while the mutual forces also point toward the opposite galaxy's centre of mass and hence provide a (smaller) positive torque, explaining the counterintuitive sign change. 

The torque caused by the self-friction of the disrupting satellite is roughly the same in the first three columns of Figure~\ref{fig:torques}, and absent by construction in the last panel (point-like satellite). On the other hand, the torque caused by the self-force of the host galaxy is present only in the \LH cases with a single satellite that creates an asymmetric perturbation (first and last columns). The physical mechanism of this torque is the internal deformation of the host: the dense central parts swing toward the pericentre of the satellite's trajectory (middle panel of Figure~\ref{fig:snapshots_gamma2}), and then are pulled back by the gravity of the outer parts, which were too slow to react (as shown by the solid line in the right panel of that figure). Both these self-forces are directed roughly perpendicularly to the line connecting the host and the satellite, and thus primarily affect the angular momentum rather than orbital energy. Ultimately, both effects are required for a noticeable radialization of the orbit.

Although we have discussed in detail only one case ($\gamma=2$, $\eta=0.3$), the time evolution of torques is qualitatively similar for other values of parameters (Figure~\ref{fig:torques_snapshots_gamma13}). However, we recall that the orbits always circularize in the steep cusp ($\gamma=3$) case. This can be explained by its very centrally concentrated density profile, so that both galaxies do not develop significant dipole perturbations (much less deformation is seen in the bottom right panel of that figure) and are more closely approximated as point masses (although the satellite still loses mass), and as we have seen, this suppresses the radialization.

%%%%%%%%%%%%%%%%%%%%%
\section{Conclusions}   \label{sec:conclusion}

%%%%%%%%%%%%%%
\begin{figure}
\includegraphics{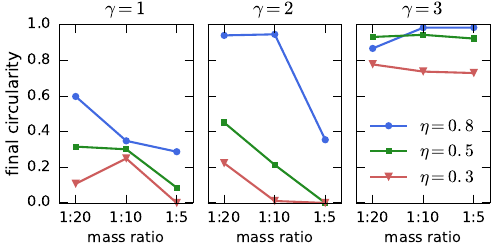}
\caption{Orbital circularity at the time of satellite's near-disruption (when its bound mass dropped to $10^{-3}$ of initial mass) as a function of mass ratio $q$ for the simulations shown in Figures~\ref{fig:evolution_gamma1}--\ref{fig:evolution_gamma3} (essentially the last point in the middle row of each figure). Three panels show different cusp slopes ($\gamma=1,2,3$), and different colours show initial circularities of 0.3 (red), 0.5 (green) and 0.8 (blue). The overall trend is that the radialization is stronger for smaller $\gamma$, lower initial circularity $\eta$, and higher $q$. Models with $\gamma=3$ circularize in all cases (the model with $q=1:20$ and $\eta=0.8$ was terminated well before the disruption and did not reach full circularization).
}  \label{fig:summary}
\end{figure}
%%%%%%%%%%%%

We have explored the behaviour of satellite orbits in merger simulations, focusing on the evolution of eccentricity. Figure~\ref{fig:summary} summarizes our findings and illustrates the following factors that contribute to the radialization of orbits:
\begin{itemize}
\item high satellite mass (mass ratio $q \gtrsim 0.1$);
\item low initial circularity ($\eta \lesssim 0.5$);
\item shallow density profile of the host galaxy (density slope $\gamma\le 2$): models with $\gamma=1$ radialize even for lower $q$ or higher $\eta$, while models with $\gamma=3$ always circularize.
\end{itemize}
We use various devices to selectively disable certain physical ingredients that might be responsible for the radialization. By performing control experiments in different setups and examining the torques acting on both galaxies from their opponents and from their own asymmetric mass distribution, we find that the increase of eccentricity is caused by a combination of several effects:
\begin{itemize}
\item Self-friction created by the debris tidally stripped from the satellite after each pericentre passage. The self-friction force is not very large in absolute terms, in agreement with \citet{Fujii2006}, \citet{Miller2020}, but its direction creates a significant, even if short-lived, negative torque (i.e., decreases the orbital angular momentum). This effect is absent by construction in simulations with point-like satellites, which do not exhibit radialization, and is less prominent for lower-eccentricity orbits, since the tidal shock during the pericentre passage is weaker. Earlier studies from 1980s and 1990s mostly considered point-like satellites and circular orbits, thus have missed this effect.
\item The motion of the host galaxy's densest central region toward the pericentre of the satellite orbit displaces the centre of density relative to the centre of mass (this offset is actually observed in interacting galaxy pairs, e.g., \citealt{Davoust1988}). As a result, shortly after the pericentre passage, the restoring force from the outer parts of the host acting on its centre also creates a significant negative torque. This torque is actually larger than the direct torque from the host on the satellite itself, caused by the local density wake in the host (the only ingredient in the CDF picture). When this dipole perturbation in the host is artificially suppressed, either by a symmetric placement of two satellites or by replacing the live host potential by the initial analytic profile, the orbital angular momentum does not drop as much. This settles the question about the local vs.\ global nature of the friction force raised in earlier studies (e.g., \citealt{Zaritsky1988}, \citealt{Hernquist1989}): at least in the case of sufficiently massive satellites, the reflex motion of the host centre \textit{and} the additional dipole perturbation appear to dominate the host response, in agreement with \citet{Tamfal2021}. One may argue whether the first factor (reflex motion) should count as part of the global perturbation (as advocated by \citealt{Weinberg1989}) or not (as viewed by \citealt{Prugniel1992} and other authors); however, it is the second factor -- the offset between the centres of mass and density -- that truly matters for the radialization.
\end{itemize}

As Figure~\ref{fig:summary} demonstrates, the radialization is always stronger for higher satellite masses, since these perturb the host galaxy to a larger degree, and their debris exert larger retarding torque. Likewise, high initial eccentricity creates stronger shocks and larger displacements of the host centre relative to its outer parts due to a shorter duration of pericentre passage. Finally, the observed unconditional circularization of $\gamma=3$ models is likely explained by their more ``stiff'', non-deforming nature.

Both the host and the satellite self-gravity appears to play an important role in the radialization, and both effects peak shortly after the pericentre passage. By contrast, in the classical CDF picture we would expect the retarding torque to be applied roughly symmetrically before and after the pericentre. However, in $N$-body simulations, we see that the CDF-like torque by the host on the satellite actually changes sign around pericentre; this counterintuitive behavior is again driven by the offset between the host centre of mass (where the host-induced satellite acceleration is pointing to) and its centre of density (which is the natural reference point for measuring the angular momentum and the torque). In the end, it seems that while CDF can successfully predict the evolution of orbital size (i.e., energy loss), it fails to match the evolution of angular momentum, which comes as no surprise since it is driven by global perturbations. It remains to be seen if a suitable modification of CDF can be designed to rectify this deficiency.

\section*{Acknowledgements}
We thank the referee for drawing our attention to the unpublished work by Amorisco presented at the conference "Stellar halos across the cosmos", which also identified the key role of the self-force of both tidally deformed galaxies in the radialization process. EV acknowledges support from STFC via the Consolidated grant to the Institute of Astronomy.

%%%%%%%%%%%%%%%%%%%%%%%%%

\end{document}